\documentclass{article}

\usepackage{arxiv}

\usepackage[utf8]{inputenc} % allow utf-8 input
\usepackage[T1]{fontenc}    % use 8-bit T1 fonts
\usepackage{hyperref}       % hyperlinks
\usepackage{url}            % simple URL typesetting
\usepackage{booktabs}       % professional-quality tables
\usepackage{amsfonts}       % blackboard math symbols
\usepackage{nicefrac}       % compact symbols for 1/2, etc.
\usepackage{microtype}      % microtypography
\usepackage{lipsum}
\usepackage{graphicx}
\usepackage{algorithm}
\usepackage{algorithmic}
\usepackage{multirow}
\title{Fake News Detection through Graph Comment Advanced Learning}

\author{
  Hao~Liao \\
  Shenzhen University\\
  Shenzhen, China\\
  \texttt{haoliao@szu.edu.cn} \\
   \And
 Qixin~Liu \\
  Shenzhen University\\
  Shenzhen, China\\
  \texttt{1810273008@email.szu.edu.cn} \\
  \And
 Kai~Shu \\
  Illinois Institute of Technology\\
  Chicago, Illinois, United States\\
  \texttt{kshu@iit.edu} \\
  \And
 Xing~Xie \\
  Microsoft Research Asia\\
  Beijing, China\\
  \texttt{xingx@microsoft.com} \\
}

\begin{document}
\maketitle

\begin{abstract}
\lipsum[1]
Disinformation has long been regarded as a severe social problem, where fake news is one of the most representative issues. What is worse, today's highly developed social media makes fake news widely spread at incredible speed, bringing in substantial harm to various aspects of human life. Yet, the popularity of social media also provides opportunities to better detect fake news. Unlike conventional means which merely focus on either content or user comments, effective collaboration of heterogeneous social media information, including content and context factors of news, users' comments and the engagement of social media with users, will hopefully give rise to better detection of fake news.

Motivated by the above observations, a novel detection framework, namely graph comment-user advanced learning framework (GCAL) is proposed in this paper. User-comment information is crucial but not well studied in fake news detection. Thus, we model user-comment context through network representation learning based on heterogeneous graph neural network. We conduct experiments on two real-world datasets, which demonstrate that the proposed joint model outperforms 8 state-of-the-art baseline methods for fake news detection (at least 4\% in Accuracy, 7\% in Recall and 5\% in F1). Moreover, the proposed method is also explainable.
\end{abstract}

% keywords can be removed

\section{Introduction}
Fake news refers to false and often inflammatory information disseminated under the guise of news reports. In this era of rapid development of social networks, fake news is common on social networks. Many bloggers choose to publish completely fake news because a large percentage of readers tend to share content without reading carefully, which can attract attention and profit for these bloggers~\cite{yang2019unsupervised,grinberg2019fake,zhang2020overview,jin2017multimodal}.
According to the daily bulletin of the World Health Organization on February 2nd, the outbreak of the COVID-19 has been accompanied by a massive 'infodemic', which means an over-abundant of information - some accurate and some not~\footnote{~\url{https://www.who.int/docs/default-source/coronaviruse/situation-reports/20200202-sitrep-13-ncov-v3.pdf?sfvrsn=195f4010_6/}}. Information flooding makes it difficult for people to distinguish fake information from facts, which is easy to cause social disorder and disturb the control of the epidemic. In 2018, the European Commission released a report "Fake News and Disinformation Online" in Eurobarometer~\cite{ZA6934}. The results of the report show that widespread fake news has harmful social effects, slowing people's trust in social media. This indicates respondents are less trusting of online news and information than traditional sources. Moreover, against the background that fake news is becoming a global public hazard, more and more countries begin to reflect and strengthen measures to control fake news and ensure the spread of accurate information in society. Therefore, social media is responsible for preventing the spread of fake news in order to strengthen trust in the entire news ecosystem and it is important to detect fake news in social media~\cite{shimizu20202019,shu2019defend,boididou2017learning,zhang2019multi,papanastasiou2020fake}.
%
%,

 %37$\%$ respondents encountered fake news almost every day, and another 31$\%$ said that it happened at least once a week. On the other hand, a majority of respondents fully trust or tend to trust the news and information they received by radio (70$\%$), television (66$\%$) and print media (63$\%$). But 47$\%$ respondents trusted online newspapers and magazines, even less trusted video sites(27$\%$) and social media and apps (26$\%$).

% Collins Dictionary published ten annual hot words in 2017, and the word "fake news" was at the top of the list~\footnote{~\url{https://www.collinsdictionary.com/word-lovers-blog/new/collins-2017-word-of-the-year-shortlist,396,HCB.html}}.
%  In 2018, for example, French President Emmanuel Macron promised in his New Year's greetings to the media that the government would submit a new law to parliament to combat the spread of fake news during elections~\footnote{~\url{https://www.theguardian.com/world/2018/jan/03/emmanuel-macron-ban-fake-news-french-president}}.
However, the advent of social media means that true and fake news is presented in a similar way, sometimes hard to distinguish. First, in order to attract readers, the content of fake news is mixed with true and false information. Second, social media data is massive and various, e.g., a large number of anonymous users transmit noisy information. The recent research about fake news detection on deep learning algorithms has achieved notable success~\cite{girgis2018deep,liu2018early}, which utilize various news features on social media, such as text content, user characteristics and user comments. However, context learning for fake news has not been maximized. Specifically, first of all, it is not enough to get accurate fake news prediction only according to text content~\cite{popat2017assessing,gupta2014tweetcred}, because context on social media is usually short and fragmented. Secondly, another research on fake news detection is based on analyzing features of first news spreaders, while ignoring the opinions of subsequent user comments~\cite{yang2012automatic,castillo2011information}.

In order to deal with the above-mentioned limitations of the existing methods, in this paper, we propose a novel method to study the challenge of fake news detection. First, we study words context and sentences context to model news content representation. In addition, we obtain potential nodes information and rich neighborhood information among nodes by establishing a heterogeneous graph between news users and news comments. Therefore, we build a graph comment-user advanced learning framework through 1) A heterogeneous graph neural network can obtain the representation of users and comments, which better utilizes the user-comment information on social media to understand the user-comment context. The example illustration is shown in Fig. 1. In this example, the news content is about "preparing to arrest sanctuary city leaders", which is related to words such as "anarchy" and "rebellious usurpers" in the comments. Therefore, we can use the relevant content in comments to help detect the authenticity of a piece of news. 2) A text representation module based on the unified pre-trained language model~\cite{dong2019unified} for news content learning, which uses the self-attention learning mechanism to capture the longer distance dependency between sentences more efficiently and deeply. The contributions of this paper are summarized as follows:
\begin{itemize}
\item{We propose a novel GNN-based model named Graph Comment-User Advanced Learning (GCAL) for fake news detection, that jointly learn explicit and implicit contextualized representation effectively in news content and user-comment interactions.}

\item{We provide a novel perspective for fake news detection to derive information from user comments by a heterogeneous graph network construction that captures both structure and content heterogeneous information.}
\item{We conduct extensive experiments on two real-world datasets, and our results demonstrate that the superior effectiveness of GCAL over state-of-the-art graph-based and other new methods for fake news detection and explainable sentences selection.}
\end{itemize}
% The proposed model identifies the contextualized representation of news content through the pre-trained language model, which can efficiently deal with natural language understanding problems.%

\section{Related works}
\subsection{Fake news detection}
We have all encountered fake news, but it is difficult to identify fake news from massive news. This question leads researchers to explore various ways for evaluating the authenticity of a piece of news. In recent years, as an emerging research direction, fake news detection is still in a slow development stage.

\textbf{Content-based methods}: Knowledge plays a crucial role in the deep excavation of news content. The method based on knowledge graph compares the extracted knowledge from news to evaluate the authenticity of news content~\cite{wang2017liar}. The New news is happening every day, which can bring new knowledge to promote fake news detection. Reporters have the same writing style as writers. News style is the ability to express facts and spread information, and it can also be the result of the work practice of journalists. Currently, the method of news style is devoted to capturing fake information from the content~\cite{ruchansky2017csi}.
For example,~\cite{pennebaker2015development} proposed the text analysis tool (LIWC) based on linguistic inquiry and word count, and in the same year, the sentence relevance research based on rhetorical structure theory (RST) was proposed~\cite{rubin2015truth}. In 2018, the convolutional neural network (CNN)~\cite{wang2018eann,volkova2017separating}  can also automatically learn content features to detect rumors.

\textbf{Social network propagation-based methods}:
On social media, the spreading power of truth cannot compete with rumors. In 2018, The Atlantic investigated 3 million tweeters published by 126 thousand users on social media from 2006 to 2016. In each evaluation, fake news spread faster and wider than the truth~\footnote{~\url{https://www.theatlantic.com/technology/archive/2018/03/largest-study-ever-fake-news-mit-twitter/555104/}}. Based on this phenomenon, the evaluation method of news credibility based on homogeneous network was proposed by~\cite{jin2016news,yang2019unsupervised,zhang2018fake},
%,
and the network structure method based on news propagation on social media has been proposed by~\cite{wang2015detecting}.

\textbf{Explainable learning-based methods}:
Explainability can help users better understand the behavior of the model, explain the results of the model, and improve the trust and transparency of the model~\cite{dovsilovic2018explainable}. There are usually two types of explainability: integrating the explanation module into the model construction or using a new model to explain the results of the current model. In 2019,~\cite{shu2019defend} proposed a method which could explain fake news detection, and it used a sentence-comment joint attention mechanism to capture the semantic consistency of news sentences and news comments.

\begin{figure}[t!]
\centering
\includegraphics[width=12cm, height=9cm]{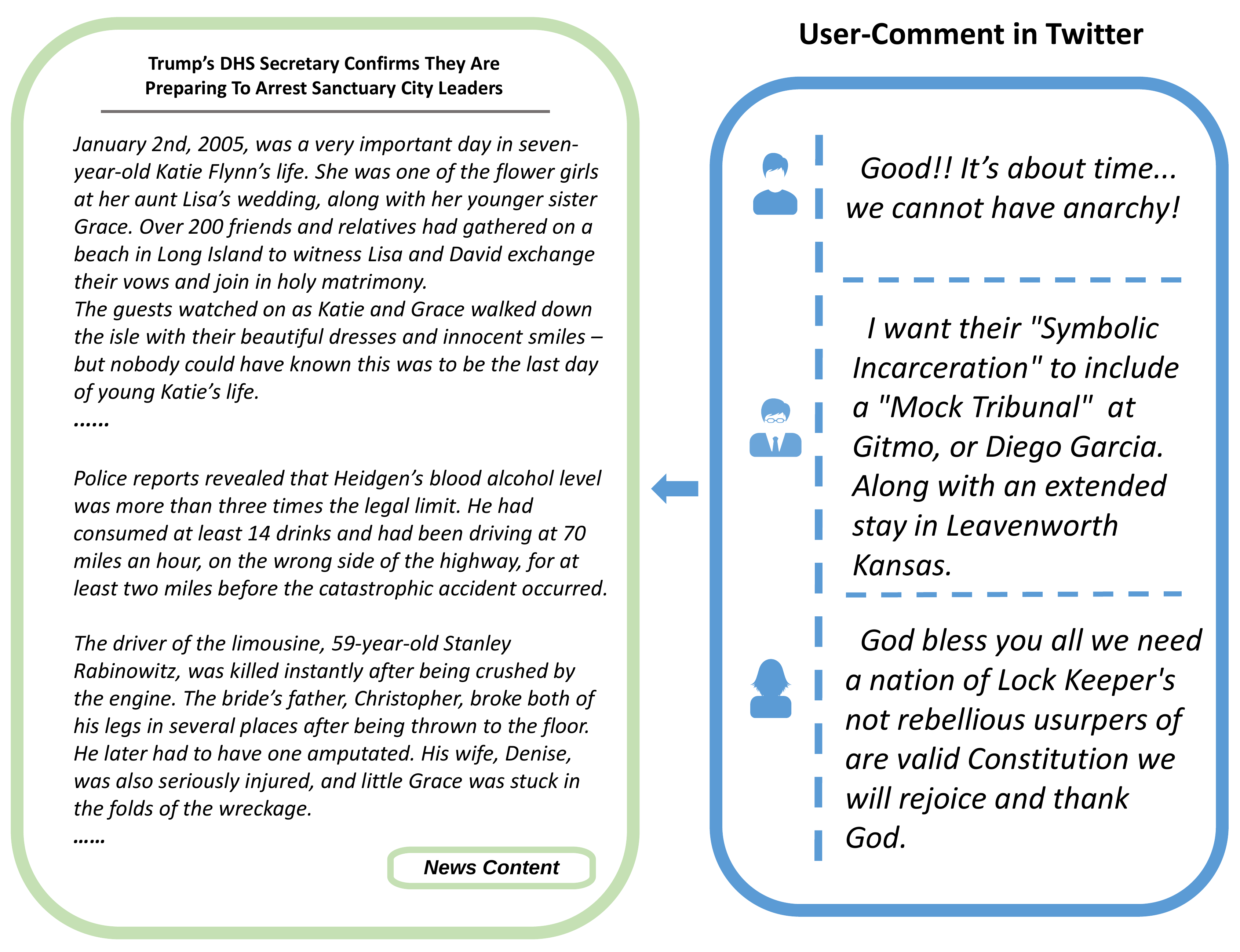}
\caption{Example of social media and news content interaction, which consists of news content and user comments. Some important user comments are highly correlated with sentences in the news content.}
\label{fig:model_example}
\end{figure}

\begin{figure*}[!htbp]
  \includegraphics[width=\textwidth]{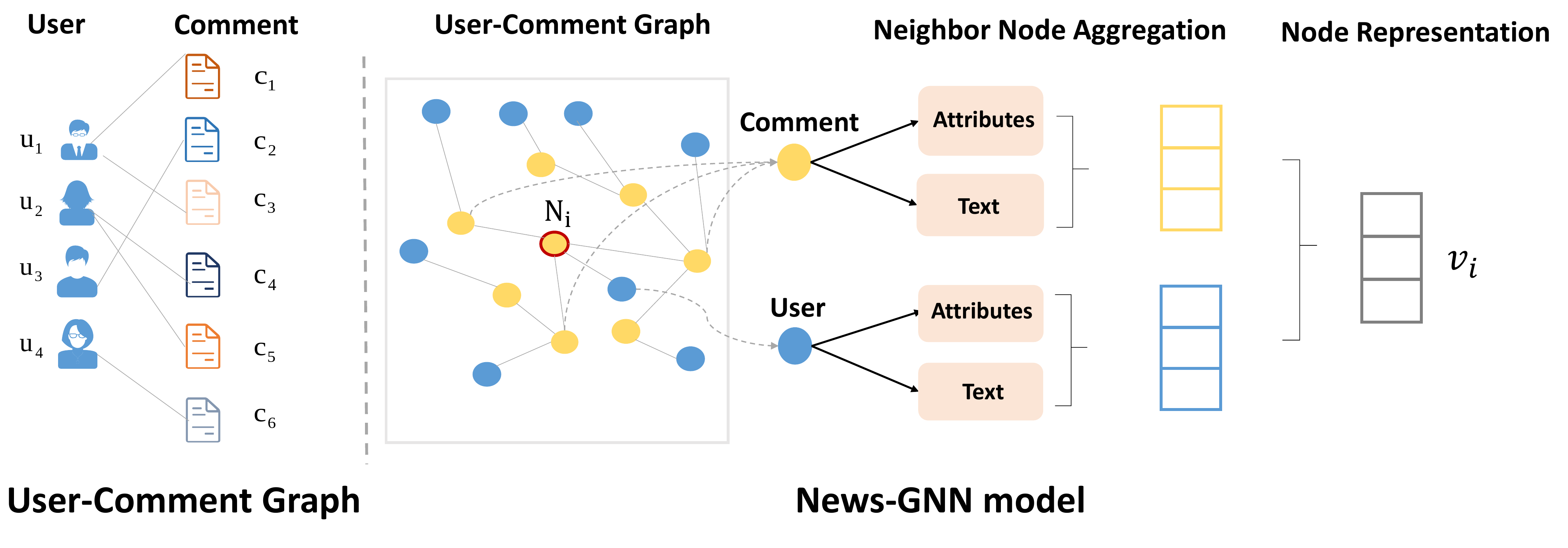}
  \caption{User-comment Graph and News-GNN model: the node representation $v_i$ of $N_i$ is integrated by aggregating its two types neighbors features.}
  \label{fig:structure1}
\end{figure*}

\subsection{Text representation based on neural networks}
Nowadays, with the gradual development of neural networks, more and more text representation models have been proposed, and efficient text representation models are in the top priority in text classification tasks~\cite{liu2005text}. It is divided into the following three categories:

\textbf{Word vector-based methods}: Word2Vec~\cite{mikolov2013efficient,zhang2017active}
%,
algorithm includes CBOW and Skip-gram models. CBOW learns word vectors in the prediction from the context to a target word, skip-gram learns word vectors in the prediction from the target word to the context. Later, Mikolov et al. extended the Word2Vec algorithm and proposed the Doc2vec~\cite{le2014distributed} algorithm.

\textbf{Recurrent neural network$/$ Convolutional neural network-based methods}: Text-CNN~\cite{kim2014convolutional} method is based on convolutional neural network, which inputs word vector matrix into convolution layer, pooling layer and full connection layer, and finally obtaining important local features of the text. Recurrent neural network is good well in capturing long-distance information, which can solve the shortage of CNN. Text-RNN~\cite{yang2016hierarchical} uses LSTM model to obtain sentence representation.
%However, because convolutional neural network doesn't consider the information of the words order, the relevant information between long-distance words will be lost.%

\textbf{Attention mechanism-based methods}: In the natural language machine translation task, attention mechanism is proposed to filter a large amount of irrelevant information, which is an effective method of information selection~\cite{mnih2014recurrent,bahdanau2015neural}. Text representation methods based on attention mechanism include hierarchical attention and self-attention. Hierarchical attention means that word encoder obtains word vector representation, and then inputs sentence vectors composed of all word vectors into sentence encoder to obtain text representation~\cite{ying2018sequential}. Self-Attention uses bidirectional LSTM to obtain the sentence representation~\cite{shaw2018self}. The Transformer model~\cite{vaswani2017attention} and BERT~\cite{DevlinCLT19} model all use the self-attention method.

% \subsection{Network Embedding}
% Network embedding transforms complex network information into structured multi-dimensional features by transforming the network structure into multi-dimensional vectors. These multi-dimensional features are convenient for algorithm applications in machine learning. Metapath2vec~\cite{dong2017metapath2vec} builds heterogeneous neighbors of nodes through the metapath-based random walk, captures connections between different types of nodes, and obtains the heterogeneous network representation. Through graph structure closeness and text semantic correlation, SHNE~\cite{zhang2019shne} conducts jointly optimization of node embedding for learning heterogeneous graph.
\subsection{Graph neural networks}
%Graph neural network is a neural network that learns directly on graph structure, and it is very popular in social networks, knowledge graph and recommendation systems.
Graph neural network captures the interrelation between nodes by mapping the data of non-euclidean space~\cite{micheli2009neural,scarselli2008graph}. GAT~\cite{VelickovicCCRLB18} model uses attention mechanism to distribute different weights for the features of adjacent nodes, and learns the feature representation of nodes. GraphSage~\cite{hamilton2017inductive} model uses LSTM to aggregate the features of adjacent nodes of each node to represent this node, which can generate vector representation for invisible nodes. HAN~\cite{yang2016hierarchical} model uses hierarchical attention mechanism on heterogeneous graph to learn different types of connection information between nodes. HetGNN~\cite{zhang2019heterogeneous} model aggregates different types of nodes to deal with various graph mining tasks. Network embedding transforms complex network information into structured multi-dimensional features by transforming the network structure into multi-dimensional vectors. Metapath2vec~\cite{dong2017metapath2vec} builds heterogeneous neighbors of nodes through the metapath-based random walk, captures connections between different types of nodes, and obtains the heterogeneous network representation. Through graph structure closeness and text semantic correlation, SHNE~\cite{zhang2019shne} conducts jointly optimization of node embedding for learning heterogeneous graph.

\section{Problem Statement}
In this section, we will define the detection problem of fake news on social media. Given a news piece $A$ consisting of $N$ sentences, just as $A= \left\{s_{i}\right\}^{N}_{1}$. In the news $A$, each sentence $s_{i} = \left\{w^{i}_{1},...,w^{i}_{M_{i}}\right\}$ includes $M_i$ words. Let $C$ be the user comments set of the new $A$ and it contains $K$ comments denoted as $C= \left\{c_{i}\right\}^{K}_{1}$. Similarly, each comment $c_{i} = \left\{w^{i}_{1},...,w^{i}_{P_{i}}\right\}$ composed of $P_{i}$ words. Among all comments, each comment corresponds to each user. Denote the all users expressing their comments as $U=\left\{u_{i}\right\}^{T}_{1}$, where $u_{i}$ is the user corresponding to the comment $c_{i}$. Following previous researches, we define fake news detection as a binary-class classification tasks. Given a news $A$, all users and comments information about the news $A$, the model proposed in this paper aims at learning a binary-class classification function $f: f(A, C, U) \rightarrow (\hat{y})$ to maximize the accuracy in news classification task.

\section{The Proposed Framework - GCAL}
\begin{figure*}[t!]
\centering
\includegraphics[scale=0.35]{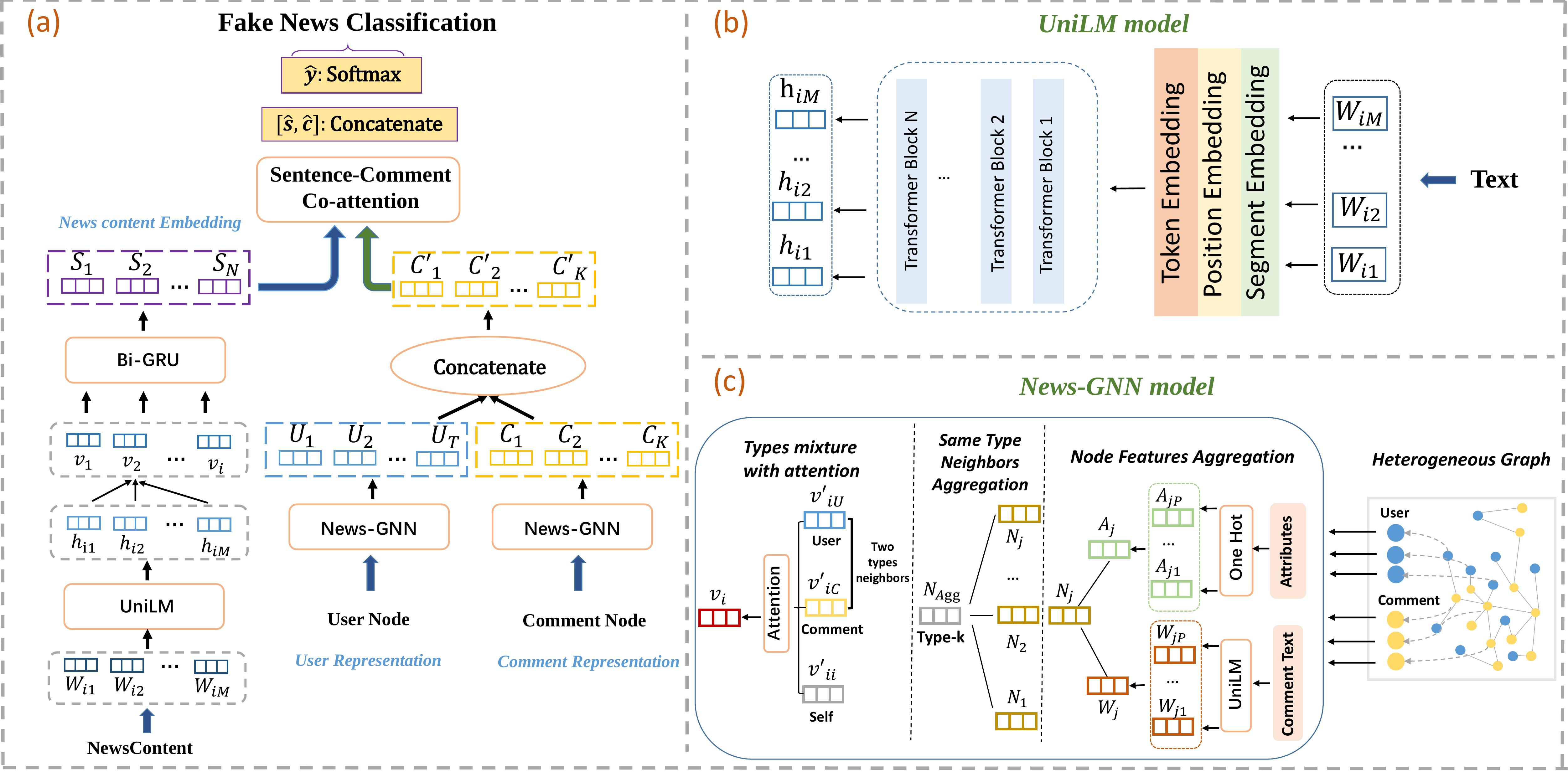}
\caption{GCAL architecture: (a) The proposed framework GCAL. (b) News-GNN model components.}
\label{fig:structure1}
\end{figure*}

In this section, we propose a novel fake news detection framework called \textbf{\underline{G}raph \textbf{\underline{C}}omment-user \textbf{\underline{A}}dvanced \textbf{\underline{L}}earning} (GCAL). GCAL follows a hierarchical attention structure, which includes the following two parts:

\textbf{Part 1: Attentional learning based on text representation.} In this part, we construct the attention learning mechanism based on a unified pre-trained language model (UniLM)~\cite{dong2019unified} to explore the relation between news sentences.

\textbf{Part 2: Network representation learning based on heterogeneous graph neural network.} As shown in Fig. 2, we establish a user-comment graph neural network to learn the representation of different types of nodes, and capture structure information between nodes and comment text information of nodes. In user-comment graph, a node contains text information (e.g., the comment text) and its own attribute information (e.g., the number of user followers and friends). Furthermore, we combine user node representation and comment node representation into a user-comment representation.

Lastly, we finish the sentence-comment co-attention by news sentences representation and user-comment representation, which captures the semantic affinity of sentences and comments. Fig. 3 shows the GCAL overall architecture.

\subsection{News content representation module}
The news content is the key to detect true and fake news. The fake news often has an exaggerated language style to attract people's attention so as to spread inaccurate information~\cite{shu2019defend}. A piece of news consists of multiple sentences, and a sentence consists of multiple words. The vital content of a piece of news can be obtained through word-level and sentence-level. We propose to obtain the representation of news sentences through a natural language understanding model called UniLM~\cite{dong2019unified}.

\subsubsection{Word encoder}
Inspired by the research of UniLM, UniLM model can well deal with the problem of understanding natural language in news content. A document is divided into multiple sentences and input into the UniLM model. The UniLM model can capture the relationship between sentences by learning the context content of sentences, finally, the vector representation of each sentence in this document is obtained. This model is pre-trained and can be used in three language modeling objectives: unidirectional (reading text content from left to right or from right to left ), bidirectional (reading text content from left to right and from right to left) and sentence-to-sentence prediction. The UniLM model includes segment embedding, position embedding, token embedding, and $L$-layer Transformer modules. Because a document is split into multiple sentences, and model training is completed between one sentence pair. Segment embedding is used to distinguish two sentences in a sentence pair. Specifically, if sentence $A$ and sentence $B$ are split joint, then the segment embedding will be $[0,\dots,0, 1,\dots,1]$, where the numbers of 0 and 1 are the lengths of $A$ and $B$ respectively. There may be multiple identical words in different positions in a sentence, but the vector representation of this word should be different. Position embedding is used to distinguish the positions of words in the sentence. Token embedding converts words from a token to a vector. Transformer~\cite{vaswani2017attention} is a model that uses the attention mechanism to learn the context between words in text. UniLM model reads a sentence $s_{i}$ from word $w_{i1}$ to word $w_{iM}$ as follows:
\begin{equation}
  h_{it} = UniLM(w_{it}), t\in\left\{1,\dots,M\right\}
\end{equation}
where $h_{it}$ is a word vector representation and a sentence vector $v_{i}$ composed of $M$ words vector:
\begin{equation}
  v_{i} = \sum^{M}_{t=1}{\alpha_{it}h_{it}}
\end{equation}
where $\alpha_{it}$ represents the importance of the $t^{th}$ word in sentence $i$ and can be calculated as:
\begin{equation}
    h'_{it} = tanh(h_{it})
\end{equation}
\begin{equation}
    \alpha_{it} = \frac{exp(h'_{it} h'^{T}_{W})}{\sum^{M}_{k=1}exp(h'_{ik} h'^{T}_{W})}
\end{equation}
where $h'_{W}$ is the weight parameters.

\subsubsection{Sentence encoder}
We use the recurrent neural network with GRU units to encode news sentences. The unidirectional GRU algorithm can only capture features between the current sentence and next sentence. Here we use bidirectional GRU, which can capture features between previous and next sentences. The bidirectional GRU model contains the forward GRU $\overleftarrow{f}$ and the backward GRU $\overrightarrow{f}$. We utilize a bidirectional GRU to encode sentences:
\begin{equation}
    \overleftarrow{S_i} = \overleftarrow{GRU}(v_{i}),i \in
    \left\{1,\dots,N\right\}
\end{equation}
\begin{equation}
    \overrightarrow{S_i} = \overrightarrow{GRU}(v_{i}),i \in \left\{N,\dots,1\right\}
\end{equation}
The obtained sentence vector $S_{i}$ is composed of the combination of the forward hidden layer and the backward hidden layer, i.e., $S_{i} = [\overleftarrow{S_{i}},\overrightarrow{S_{i}}]$.

\subsection{User-comment network representation learning module}
In this section, we will introduce the representation learning of users and comment nodes in details through heterogeneous graph neural network as illustrated in Fig. 3(b). Users from different professional backgrounds and offline information have various comments on news and promote fake news detection. Some users will distinguish main body of fake news from true sentences according to their own expertise or offline information source, which may affect other users’ comments. In this process, the masses who don’t know the truth will also have a negative impact, and user identity information, such as the number of followers, friends and tweets, needs to be considered. For example, users of water navy will have fewer fans, fewer friends and more comments of the same type with low quality (low quality means that the amount of giving the thumbs-up and replying and retweeting is almost zero). Therefore, we construct user-comment heterogeneous graph to strengthen the deficiencies of only by news content and dig the effective information of news comments and users, as well as the connection between news comments and news.
%For example, users of water navy will have fewer fans, fewer friends and more comments of the same type with low quality (low quality means that the amount of giving the thumbs-up and replying and retweeting is almost zero).
First, we need to establish a heterogeneous network about users and comments, and then input this network into a graph neural network to calculate the vector representation of each user node and comment node. Graph neural network consists of three parts: node features aggregation, same type neighbors aggregation, and types mixture with attention.

\subsubsection{Constructing heterogeneous graph}
Each piece of news contains multiple comments, and each comment is posted by a user. When a user receives an tweet, he or she will often scan the previous few positive or negative comments, which will often affect the comments that the user is going to post. Therefore, we assume that a comment will be related to ten comments before its release time. Because those old comments will affect the new comment posted by a user. And the comment-comment mapping is based on above. Because this paper mainly study the influence of comments information on news and the user relationship data is missing, the relationship is not taken into account when constructing a heterogeneous graph. Before constructing the user-comment graph, we have eliminated those users and comments with invalid information to avoid noise information generated during feature extraction. The heterogeneous graph we build contains only user and comment nodes, and only user-comment and comment-comment relationship pairs. In the graph, the user-comment mapping is one-to-many, and the comment-user mapping is one-to-one and the comment-comment mapping is many-to-many.

\subsubsection{Node features aggregation}
A user or a comment node contains two kinds of feature information, which are the attributes of the node and the comment text posted. The attributes of user node include the number of followers, friends, tweets published and verification flag. The attributes of comment node include the number of giving the thumbs-up, retweets and replies. We can transform node attributes into attributes vector $A_{j}$ by one-hot method and get text representation $W_{j}$ by UniLM model. Lastly, we aggregate attributes vector $A_{j}$ and text vector $W_{j}$ with mean pooling layer for a general node features representation $N_{j}$.

\subsubsection{Same type neighbors aggregation}
Neighbors of same type have similar feature representations First, we need to obtain all neighbor nodes of the current comment node through sampling, and then use Bi-LSTM module to aggregate features representation of similar neighbors. For instance, if the representation set of all type-$k$ neighbors is $\left\{N_{1},\dots,N_{j}\right\}$, we acquire all type-$k$ neighbors representation $N_{Agg}$ by Bi-LSTM module.

\subsubsection{Types mixture with attention}
Different types of neighbor nodes have different influences on nodes in a heterogeneous graph. In the user-comment heterogeneous graph, the comment nodes have two different types of neighbor nodes. The previous network representation learning model cannot be well applied in the user-comment network, because two kinds of neighbor information of a comment node are both necessary. Therefore, we use the attention mechanism to combine the feature information of two different neighbor nodes to obtain the final node feature representation. The calculation process is as follows:
\begin{equation}
    v_{i} = \alpha_{ii}v'_{ii} + \alpha_{iC}v'_{iC} + \alpha_{iU}v'_{iU}
\end{equation}
where $v'_{iv}, v\in \left\{i,C,U\right\}$, $v'_{iv}$ is the node vector representation by above process. Similarly, $\alpha_{iv}, v\in \left\{i,C,U\right\}$, $\alpha_{iv}$ indicates the importance of different embeddings and can be calculated as:
\begin{equation}
    \alpha_{iv}=\frac{\exp \left\{LeakyReLU \left(u^{T}\left[v'_{iv},v'_{ii}\right]\right)\right\}}{\sum_{j \in \left\{i, C, U\right\}} \exp \left\{LeakyReLU \left(u^{T}\left[v'_{ij},v'_{ii}\right]\right)\right\}}
\end{equation}
where LeakyReLU is a variant of the Relu activation function. Compared with Relu, it does not cause the problem that neurons do not learn when the activation function enters the negative interval, and $u$ is the parameter of attention module.

Hence, when a node is input into the graph neural network, the feature representation of all neighbor nodes of the current node will be calculated first, and then the neighbor nodes of the same type will be aggregated, finally, the integrated neighbor nodes of different types and the current node are calculated by the mixed attention, and the ultimate node embedding of current node is obtained.

\subsection{Sentence-comment co-attention module}
Not all the sentences in a piece of fake news are fake, and some real sentences are to cover the fake information. If we only rely on the true or false sentences in the news content to judge whether the news is true or false, it is inaccurate, because each sentence has different importance in identifying fake news. News comments often reflect the authenticity of the news content, and users may give some important clues of fake news detection according to their views. However, some users may be the water navy, and their comments should be less important for fake news detection. Choosing comments related to news content is crucial to fake news detection. We set up a sentence-comment co-attention mechanism, to fully learn the semantic affinity between news sentences and user comments. The news content embedding $S = \left\{S_{1},\dots,S_{N}\right\}$ and the feature representation $C'=\left\{C'_{1},\dots,C'_{K}\right\}$ of user comments can be obtained through concatenating user representation $\left\{U_{1},\dots,U_{T}\right\}$ and comment represent $\left\{C_{1},\dots,C_{K}\right\}$. We firstly calculate a conformity matrix $F$ as:
\begin{equation}
    F = \tanh{(C'^{T} W_{I} S)}
\end{equation}
where $W_{I}$ is a weight matrix. Then we are able to transform the attention mapping of sentences and comments with calculated conformity matrix:
\begin{equation}
    H^{S} = \tanh{(W_{S} S + (W_{C'} C')F)}
\end{equation}
\begin{equation}
    H^{C'} = \tanh{(W_{C'} S + (W_{S} C')F^{T})}
\end{equation}
where $W_{C'}$ and $W_{S}$ are the weight parameters. We compute attention weights of news sentences and user comments as:
\begin{equation}
    a^{S} = softmax(W^{T}_{hs} H^{S})
\end{equation}
\begin{equation}
    a^{C'} = softmax(W^{T}_{hc'} H^{C'})
\end{equation}
where $W_{hs}$ and $W_{hc'}$ is the attention weight. Based on above attention weights, the sentences embedding and comments representation can be transformed by:
\begin{equation}
    \hat{s_{i}} = \sum^{N}_{t=1}{a^{S_{t}}_{i} S_{i}}
\end{equation}
\begin{equation}
    \hat{c_{i}} = \sum^{K}_{t=1}{a^{C'_{t}}_{i} C'_{i}}
\end{equation}
where $\hat{s_{i}}$ and $\hat{c_{i}}$ are learned from co-attention mechanism. Aggregate $\hat{s_{i}}$ and $\hat{c_{i}}$ for the news prediction:
\begin{equation}
    \hat{y} = softmax(W_{f}[\hat{s},\hat{c}] + b_{f})
\end{equation}
where $W_f$ is the weight parameter and $b_{f}$ is the bias value. $\hat{Y} \in [\hat{y_{0}},\hat{y_{1}}]$ is used as the prediction value for fake news detection. In binary classification for fake news detection, our goal is to minimize the following loss function:
\begin{equation}
    Loss = -y\log{(\hat{y_{1}})} - (1-y)\log{(1-\hat{y_{0}})}
\end{equation}
where $y \in [0,1]$ is the true news label, which represents fake news and true news repectively. We will describe the overall algorithm of GCAL in Algorithm 1.

\begin{algorithm}[tb]
\caption{GCAL}
\label{alg:1}
\leftline{\textbf{Input:} The news set $NS$; all news comments $NC$; all news users}
\leftline{$NU$; all nodes attributes $NA_{i}, i \in \left\{NC, NU\right\}$}

\leftline{\textbf{Output:}
The news prediction probability vector $\hat{Y}$}

\begin{algorithmic}[1] %[1] enables line numbers
\STATE Constructing heterogeneous graph $G = (V, E)$,

$V \in (NC, NU)$ and $E$ includes relation between users and comments;
\STATE // Comment and user nodes embedding
\FOR{$i \in \left\{NC \cup NU\right\}$}
    % \STATE $N_{i}$ = FeaturesAggr($NA_{i}$);
    % \STATE Neighbor nodes set $N_{neigh_{i}}$;
    % \FOR{$j \in N_{neigh_{i}}$}
    %     \STATE $N_{j}$ = FeaturesAggr($NA_{j}$);
    % \ENDFOR
    % \STATE $N_{agg_{C}} = NeighborsAggr(\left\{t \in (NC \cup N_{neigh_{i}}) \mid N_{t}\right\})$;
    % \STATE $N_{agg_{U}} = NeighborsAggr(\left\{t \in (NU \cup N_{neigh_{i}}) \mid N_{t}\right\})$;
    \IF{$i \in NC$}
        % \STATE $C_{i} = MixAttention(N_{i}, N_{agg_{C}}, N_{agg_{U}})$
         \STATE $C_{i} = NewsGNN(NA_{i})$
    \ELSE
        % \STATE $U_{i} = MixAttention(N_{i}, N_{agg_{C}}, N_{agg_{U}})$
        \STATE $U_{i} = NewsGNN(NA_{i})$
    \ENDIF
\ENDFOR
\FOR{$A \in NS$}
    \STATE // News content embedding.
    \STATE //Each news $A$ has $N_A$ sentences.
    \FOR{$i \in [1,N_{A}]$}
        \STATE // Each sentence $i$ has $M_{A_i}$ words.
        \FOR{$t \in [1,M_{A_{i}}]$}
            \STATE $h_{it}$ = UniLM($w_{it}$);
            \STATE Calcalate the importance $\alpha_{it}$ of $t^{th}$ word in sentence $i$;
        \ENDFOR
        \STATE $v_{i} = \sum^{M_{A_{i}}}_{t=1}{\alpha_{it}h_{it}}$;
        \STATE $S_{i} = $Bi-GRU$(v_{i})$;
    \ENDFOR
    \STATE Let all sentences representation of news $A$ as $S_{A} = \left\{S_{t},t \in [1,N_{A}]\right\}$;
    \STATE Get all users representation $U_{A}$ and comments representation $C_{A}$ about news $A$ from step3 to step9;
    \STATE Concatenate $U_{A}$ and $C_{A}$ as $C'_{A}$;
    \STATE $[\hat{s},\hat{c}]$ = Co-attention($S_{A},C'_{A}$);
    \STATE Computer news prediction $\hat{y} = softmax(\hat{s},\hat{c})$;
\ENDFOR

\STATE $\hat{Y} = \left\{A \in NS \mid \hat{y}_{A}\right\}$;
\RETURN $\hat{Y}$;
\end{algorithmic}
\end{algorithm}

\section{Experiments}
In order to identify the effectiveness of our proposed model, we designed empirical studies. In this section, we firstly introduce the dataset description and the state-of-the-art baseline methods. After that, the experiments will be conducted to evaluate the performance of the GCAL from various metrics.

\subsection{Experiments setup}
\subsubsection{Datasets}
We deploy our experiments on two datasets of fake news detection, which are collected from FakeNewsNet~\cite{shu2019beyond}. These two public datasets are Politifact and Gossipcop. Their key statistics are shown in table 1. Detailed introduction about two datasets will be described below:

\textbf{\textit{Politifact:}} In this dataset, the news are divided into real news and fake news by considering the journalists and the expert reviews on the political news on the website.

\textbf{\textit{Gossipcop:}} In this dataset, the entertainment news with rating score are collected from social media. The FakeNewNet only considers the news with rating score less than 5.

\begin{table}[t!]
\centering
\small
\caption{Data Statistics}
\setlength{\tabcolsep}{5mm}{
\label{tab::datasets}
\begin{tabular}{lrr}
\toprule[1pt]
\multicolumn{1}{l}{\textbf{Platform}} & \multicolumn{1}{c}{\textbf{PolitiFact}} & \multicolumn{1}{c}{\textbf{GossipCop}} \\ \hline
 \# Users & 36,060 & 95,319 \\
 \# Comments-Users & 73,373 & 129,710 \\
 \# True News & 152 & 1,112 \\
 \# Fake News & 237 & 862 \\
  \bottomrule[1pt]
\end{tabular}
}
\end{table}

These datasets contain news content with labels and social information on users and comments. Comments corresponding to each piece of news are acquired by the FakeNewsNet tool. Comment information includes the comment text and attributes (e.g., reply count, retweet count and liking count). User information consists of all comments text and properties (e.g., friends count, followers count, verified flag and statuses count). More detailed information can be clarified in~\cite{shu2019beyond}.

\subsubsection{Baselines}
We conducted our experiments comparing with the following methods. These methods are classified into two categories: Graph neural network methods and Text classification methods.

\textbf{\textit{Graph neural network methods}}

\begin{itemize}
\item {\textbf{HetGNN~\cite{zhang2019heterogeneous}}}: HetGNN is a heterogeneous graph neural network for various graph mining tasks by aggregating different types of nodes.
\item {\textbf{GAT~\cite{VelickovicCCRLB18}}}: GAT uses self-attention neural network to aggregate neighbors’ features for various tasks.
\item {\textbf{GSAGE~\cite{hamilton2017inductive}}}: GSAGE generates a aggregator for node embedding by sampling and collecting features from neighbors.
\end{itemize}

\textbf{\textit{Text classification methods}}

\begin{itemize}
% \item {\textbf{RST}}: RST is the abbreviation of rhetoric structure theory. The algorithm represents rhetorical relations among the words by creating a tree structure.
\item {\textbf{HAN~\cite{yang2016hierarchical}}}: HAN combines node-level attention and semantic-level attention to learn information on news content.
\item {\textbf{text-CNN~\cite{kim2014convolutional}}}: text-CNN combines the convolutional neural networks and news contents. By utilizing mutiple convolution hidden layers, it can automatically extract text features.
\item {\textbf{text-RNN~\cite{yang2016hierarchical}}}: text-RNN uses LSTM to encode text information in the last output of recurrent neural network.
\item {\textbf{TCNN-URG~\cite{qian2018neural}}}: TCNN-URG utilizes a two-level convolutional neural network and a conditional variational auto-encoder for classification.
\item {\textbf{dEFEND~\cite{shu2019defend}}}: dEFEND is a model that develop a sentence-comment co-attention sub-network to exploit both news contents and user comments to jointly capture explainable top-$k$ check-worthy sentences and user comments for fake news detection.
\end{itemize}

\subsubsection{Evaluation metrics}
In our experiments, we evaluated the results in binary-classification task by the most frequently-used metrics Accuracy, Precision, Recall, F1 and AUC~\cite{shu2017fakesurvey}.

\subsubsection{Reproducibility}
By randomly splitting the dataset into training (75$\%$) and validation (25$\%$), we acquire the set of news nodes as the target node to conduct the fake news detection. We use the same embedding dimension $d=200$ for each baseline algorithm. In our proposed GCAL framework, due to considering the time performance, we set the maximal length of news sentences 50 and maximal comment length 20 in two datasets. Differently, the learning rate is 0.0002 in Politifact but 0.0015 in Gossipcop.
% To help further understand our framework, we provide complete parameters information and sufficient code details at website~\footnote{~\url{https://anonymous.4open.science/r/670f4d3c-8360-467a-a2f9-09468a83de8e/}}.

% \usepackage{multirow}
\begin{table}[]
\caption{Performance of all models}
\centering
\label{tab:performance}
\begin{tabular}{p{80pt}p{20pt}p{20pt}p{20pt}p{20pt}|p{20pt}p{20pt}p{20pt}p{20pt}}
\toprule[0.8pt]
\multirow{2}{*}{\textbf{Method}} & \multicolumn{4}{c|}{\textbf{PolitiFact}} & \multicolumn{4}{c}{\textbf{GossipCop}} \\ \cline{2-9}
 & \textbf{Acc.} & \textbf{Pre.} & \textbf{Rec.} & $\mathbf{F_1}$ & \textbf{Acc.} & \textbf{Pre.} & \textbf{Rec.} & $\mathbf{F_1}$ \\ \hline
\textbf{HetGNN}                  & .838          & .782          & .741          & .757          & .782          & .835          & .792          & .812          \\
\textbf{GAT}                     & .748          & .753          & .670          & .699          & .724          & .741          & .788          & .763          \\
\textbf{GSAGE}                   & .735          & .681          & .660          & .669          & .679          & .621          & .653          & .636          \\
\textbf{HAN}                     & .832          & .834          & .725          & .770          & .735          & .756          & .783          & .768          \\
\textbf{text-CNN}                & .789          & .737          & .712          & .717          & .657          & .659          & .785          & .716          \\
\textbf{text-RNN}                & .808          & .774          & .719          & .740          & .717          & .750          & 0.801         & .773          \\
\textbf{TCNN-URG}                & .853          & \textbf{.926} & .691          & .791          & .703          & .712          & \textbf{.854} & .777          \\
\textbf{dEFEND}                  & .887          & .892          & .811          & .847          & .817          & .830          & .852          & .840          \\ \hline
\textbf{GCAL}                    & \textbf{.924} & .917          & \textbf{.884} & \textbf{.900} & \textbf{.828} & \textbf{.847} & .850          & \textbf{.848}\\
\bottomrule[1pt]
\end{tabular}
\end{table}

\subsection{Performance evaluation}
\subsubsection{Overall performance}
In this section, we will introduce the overall performance of our proposed GCAL framework in fake news detection. All results are acquired by the average value from performing the same process for 5 times. The results of all methods are presented in Table 2, which includes the detailed evaluation metrics values.

As can be seen from the results, our GCAL model can achieve the best performance on two datasets and we noted the following analysis. Firstly, most methods with attention mechanism tend to predict more news correctly (HetGNN, GAT, GSAGE, HAN, dEFEND and ours). It implys that attention mechanism can better capture the information in news content. Secondly, methods based on sentence-comment co-attention (dEFEND and GCAL) perform better than others, because news sentences context and comments context promote to learn additional knowledge about news content. It shows that complex deep models can achieve good performance by high-level features extraction but not best (i.e., dEFEND > TCNN-URG and GCAL > TCNN-URG). In other words, additional information about news content is helpful to improve the accuracy of fake news detection. Thirdly, according to the comparison with GNNs methods, we observed that heterogeneity of networks is powerful to obtain more hidden information through constructing a heterogeneous graph neural network (i.e., HetGNN > GAT and HetGNN > GSAGE).

In terms of Accuracy, Recall and F1 in Politifact data, the current results show that our proposed GCAL has more discriminative ability than others. In terms of Precision, we find that TCNN-URG outperforms than GCAL. After detailed analysis, we observed that TCNN-URG tends to predict more true news correctly. However, our study set out to detect fake news, so higher precision and lower recall are not inaccurate. When considering Accuracy, Precision and F1 in Gossipcop data, the results indicate that GCAL has the best performance compared with other methods although slightly lower than TCNN-URG in terms of Recall. In spite of dEFEND is one of the most helpful methods in this field and utilizes straightforward attention learning to achieve powerful ability in fake news detection. Our proposed approach outperforms than HetGNN and dEFEND from all metrics perspective, which implies the effectiveness of heterogeneous neural network learning module. For example, compared with dEFEND, GCAL improves Accuracy by 4\%, Recall by 7\% and F1 by 5\% in Politifact, as well as slightly better in Gossipcop. Our design of graph comment-user advanced learning based on user-comment graph learning and pre-trained language model based on heterogeneous graph neural network, which promotes the model to mine more potential information for news content.

\begin{figure}[t!]
\centering
\includegraphics[scale=0.6]{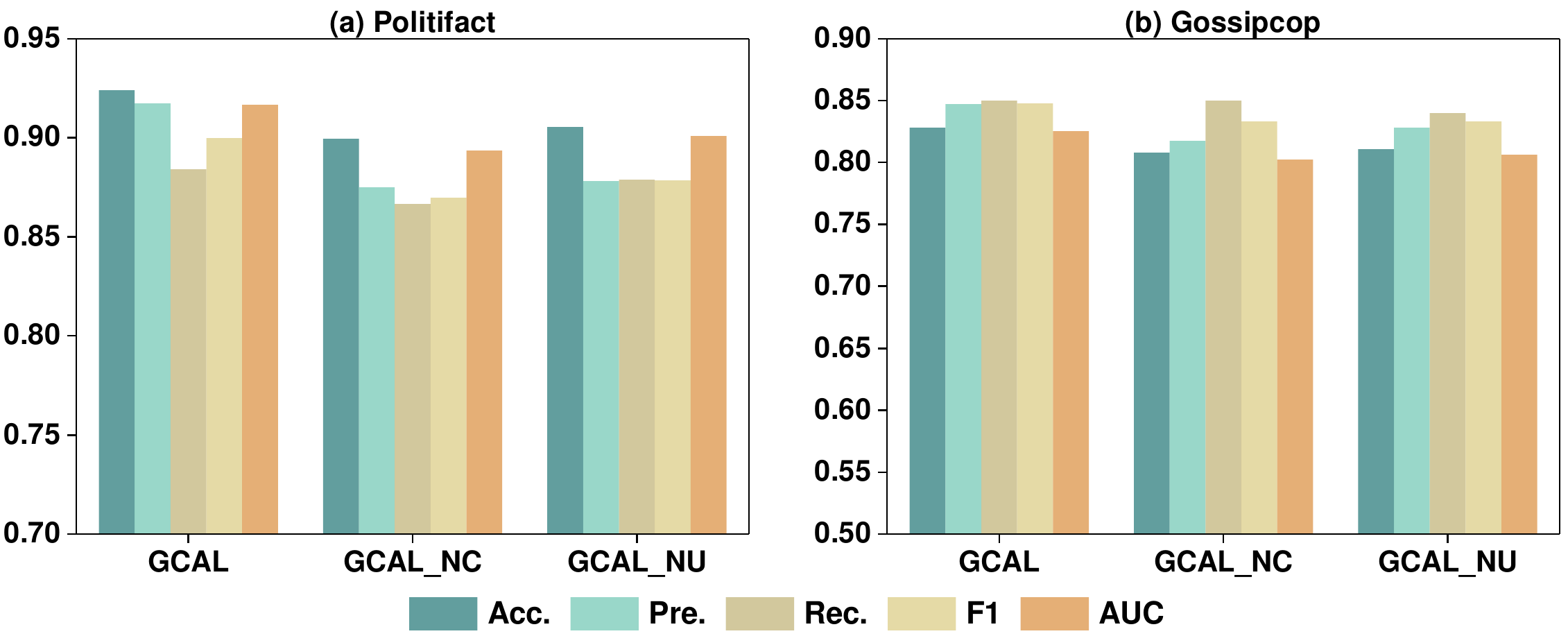}
\caption{Different metrics on Politifact and GossipCop for  GCAL, GCAL without comment representation(GCAL\_NC), GCAL without user representation(GCAL\_NU).}
\label{fig:robust}
\end{figure}

\subsubsection{Ablation analysis of GCAL} % framework
In this part, to validate the effectiveness of the framework in detail, the performance on GCAL News-GNN module in Fig. 4. We first construct a user-comment graph network and then eliminate user representation(GCAL\_NU) or comment representation (GCAL\_NC).

From Fig. 4, we can find that the GCAL has better performance than GCAL\_NC and GCAL\_NU. It proves our idea that the combination of the user attention and context feature enables GCAL to absorb features more accommodatively and perform better and stable. After the user-comment graph network construction, user nodes or comment nodes computationally aggregate two kinds of nodes context features and consequently results don't present obvious fluctuation. More importantly, the user-comment relationship and word-sentence attentional learning indeed can better help news verification.

\subsubsection{Explainability evaluation}
In this section, we will introduce the explainability on GCAL for fake news detection. During the fake news detection, we aim to learn a rank list $RS$ from all sentences. In each piece of news, the rank list consists of $k$ most explainable sentences, which play a crucial role in the task of identifying fake news. In order to analyze whether the top-$k$ explainable sentences evaluated by our proposed method are closer to those sentences that most need checking in a piece of news, we utilize ClaimBuster~\cite{hassan2017toward} to get a rank list $\tilde{RS}$ as ground truth. ClaimBuster is a tool, which has learned about 20,000 sentences from past US presidential debates for identifying worthy claims, and can calculate a reliable score between 0 and 1. Therefore, the higher the score obtained by ClaimBuster, the more consistent with the facts. Specially, we observe the performance of news sentences explainability by comparing top-$k$ rank list in news contents determined by GCAL and dEFEND, with the rank list $\tilde{RS}$ by ClaimBuster. Meanwhile, we utilize mean average precision to evaluate results, where $k$ is set as 5 and 10. During the evaluation process, another parameter $n$ is introduced to control $n$ neighboring sentences for comparison, where $n$ is set as 0 to 4. As can be seen from Fig. 5, we can get two observations. On one hand, the results show that the overall performance of finding the top-$k$ explainable sentences which are more consistent with facts in GCAL is obviously better than in dEFEND on two datasets. Accordingly, we can conclude that our proposed framework can facilitate to analyze the most need checking sentences. On another hand, due to the increase of $n$, we note that the mean average precision displays an upward trend because of the matching condition becoming easing in comparison with ground truth.

\begin{figure}[t!]
\centering
\includegraphics[scale=0.6]{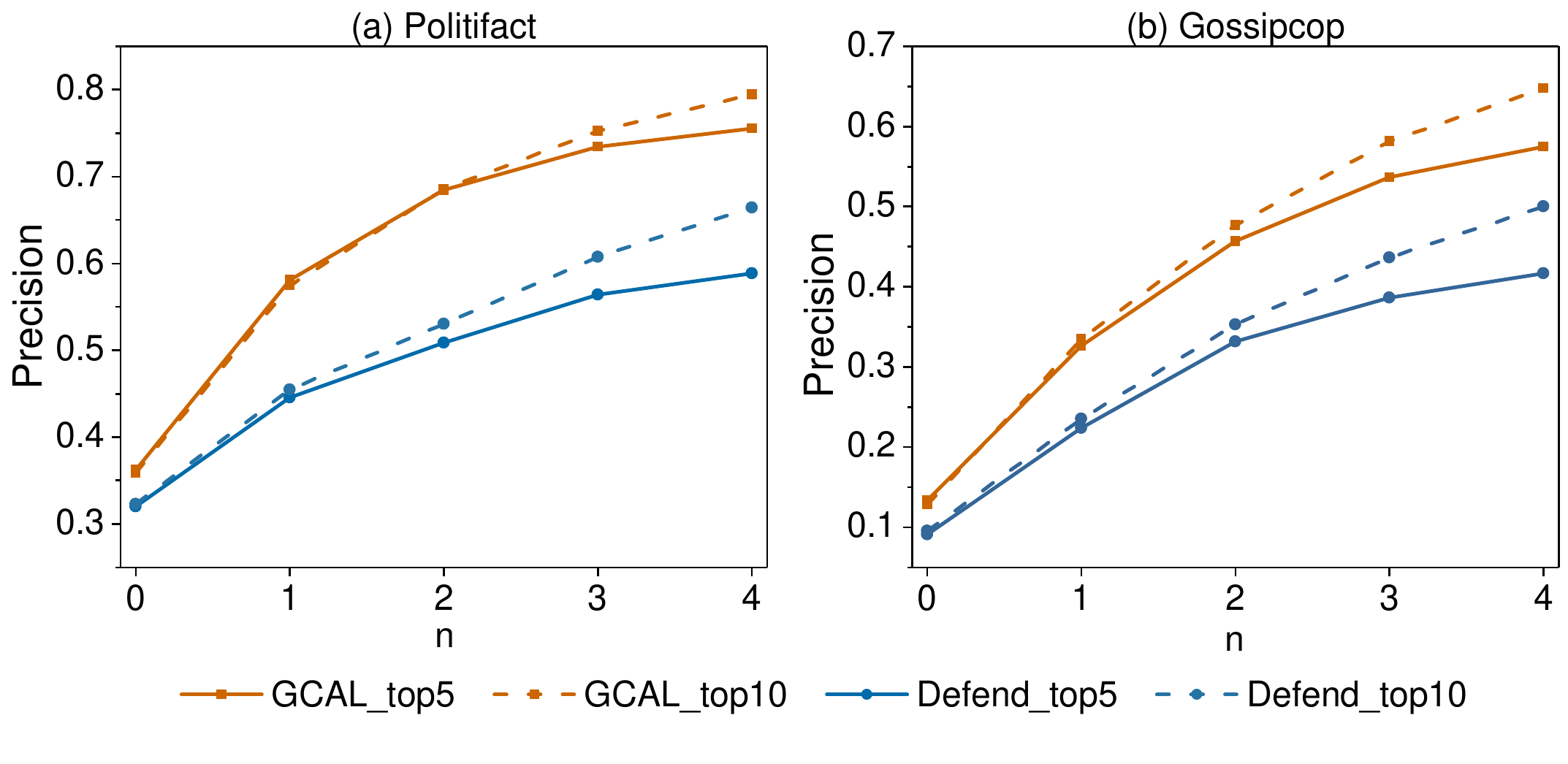}
\caption{The performance of news sentences explainability on two datasets.}
\label{fig:model_example}
\end{figure}

\section{Conclusion}
For a long time, most people were mere social media consumers. Nowadays, with the development of social media, people are actively engaged in social network interactions and the cost of producing and spreading information has become very low. Therefore, fake news detection has captured public attention. The diversified information underlying social media provides multi-perspective to facilitate fake news detection. Except for news content verification, it is worthwhile mentioning that how to make full use of heterogeneous information to mine the potential relations between news users and comments is also crucial. In this paper, we propose a GCAL framework, which utilizes heterogeneous information from news comments and news users to study the verification of fake news. In addition, the framework can also be used for the explainable sentences discovery. Experiments on two real-world datasets confirm that our proposed framework outperforms state-of-the-art methods, which suggests that it can effectively verify a piece of news on social media. We still can continue to learn more explicit explanation generation of GNN driven method for Fake news detection in the future since Graph always provides important link to the understanding.

In the future, using news spreading path can be considered as the new type of link in the graph. First, incorporating comments from authoritative users to select explainable comments for fake news verification. Second, exploring how to utilize semi-supervised learning into our model to detect massive unlabeled news on different social platform.

\newpage
\bibliographystyle{unsrt}

% Comment out this section when you \bibliography{references} is enabled.
\bibliography{references}
% George Kour and Raid Saabne.
% \newblock Real-time segmentation of on-line handwritten arabic script.
% \newblock In {\em Frontiers in Handwriting Recognition (ICFHR), 2014 14th
%   International Conference on}, pages 417--422. IEEE, 2014.

% George Kour and Raid Saabne.
% \newblock Fast classification of handwritten on-line arabic characters.
% \newblock In {\em Soft Computing and Pattern Recognition (SoCPaR), 2014 6th
%   International Conference of}, pages 312--318. IEEE, 2014.

% \bibitem{hadash2018estimate}
% Guy Hadash, Einat Kermany, Boaz Carmeli, Ofer Lavi, George Kour, and Alon
% \newblock Estimate and replace: A novel approach to integrating deep neural
%   networks with existing applications.
% \newblock {\em arXiv preprint arXiv:1804.09028}, 2018.

% \end{thebibliography}

\end{document}